\documentclass{article}
\usepackage{amsmath}
\usepackage{graphicx}
\def\g{\gamma}

\def\e{\eta}

\def\om{\omega}
\def\Om{\Omega}
\def\rh{\rho}

\def\b{\beta}

\def\pdellx'{\frac{\partial}{\partial x'}}
\def\pdellw'{\frac{\partial}{\partial w'}}
\newcommand{\be}{\begin{equation}}
\newcommand{\ee}{\end{equation}}
\def\bed{\begin{displaymath}}
\def\eed{\end{displaymath}}
\def\bea{\begin{eqnarray}}
\def\eea{\end{eqncrray}}
\def\[{$$}
\def\]{$$}
\begin{document}
 \makeatletter
% \renewcommand\theequation{{5.}\@arabic\c@equation} 
% \renewcommand\theequation{\thechapter.\@arabic\c@section} % for (1.1)keanother
%\makeatother
% \renewcommand\thesection{{2.}\@arabic\c@section} 
% \renewcommand\thesection{{5-}\@arabic\c@section} 
 %HahahahahaI found it. TMD
% \renewcommand\theequation{\thesection.\@arabic\c@equation} 
%\makeatother
%\noindent 
%\begin{document}
\title{\Large  Thim's Experiment and Exact \\ 
Rotational Space-Time Transformations  } 
%\end{center}
\author{
Leonardo Hsu\footnote{E-mail: lhsu@umn.edu}\\
Department of Postsecondary Teaching and Learning, \\
University of Minnesota, Minneapolis, Minnesota 55414,\\ 
and\\
 Jong-Ping Hsu\footnote{E-mail:jhsu@umassd.edu}\\
 Department of Physics, \\
 University of Massachusetts Dartmouth,\\ 
North Dartmouth, MA 02747\\ 
}
%\begin{center} 

%Transformations for rotating frames.....
\bigskip
\maketitle
{\small  Abstract--- Thim measured the transverse Doppler shift using a system consisting of a stationary antenna and pickup, in addition to a number of intermediate antennas mounted on the rim of a rotating disk.  No such shift was detected, although the experiment should have had enough sensitivity to measure it, as predicted by the Lorentz transformations.  However, using the Lorentz transformations to analyze the results of experiments involving circular motion, while commonly done, is inappropriate because such an analysis involves non-inertial frames, which are outside the range of validity of special relativity.  In this paper, we re-analyze Thim's experiment using exact rotational space-time transformations, finding that his null result is consistent with theoretical predictions.
\bigskip

%{\bf Index terms---Thim's experiment, transverse frequency shift, rotational space-time transformation, limiting %Lorentz-Poincar\'e invariance, absolute slow down of orbiting clocks.}} 

\bigskip

\section{Introduction}
In the paper `Absence of the relativistic transverse Doppler shift at microwave frequencies,'\cite{1}  which was followed by a comment and rebuttal \cite{2,3}, Thim describes an experiment in which microwave signals from a stationary source are received and retransmitted by two sets of antennas mounted on counter-rotating disks, and finally received by a stationary antenna. Although the Lorentz transformations of special relativity predict that the final signal received by the stationary source should exhibit a transverse Doppler shift compared to the original signal emitted by the source, none was measured, even though such a shift should have been easily within the detection sensitivity of the apparatus. The difficulty here, however, is that two reference frames that are rotating with respect to one another cannot both be inertial frames and thus special relativity and the Lorentz transformation are not applicable to this situation. 

While it is true that the Lorentz transformation has been used to analyze experiments involving rotational or orbital motion and that in some cases, the theoretical predictions are consistent with experimental results (for example in the case of calculating the lifetime dilation of unstable particles moving in a circular storage ring\cite{4}) the fact remains that, rigorously speaking, such applications are inappropriate. If the relative motion between two reference frames is rotational, then objects moving with constant velocity in one frame have non-zero accelerations as viewed from the other frame. Thus, at least one of the two frames must be non-inertial, a situation out of the purview of special relativity. In Pellegrini and Swift's analysis of the Wilson experiment\cite{5}, they demonstrate that the exact rotational transformations cannot locally be replaced by the Lorentz transformations. One major difficulty is that there does not yet exist a set of exactly correct transformations for rotating reference frames that is widely accepted and used. Nevertheless, the preceding arguments suffice to demonstrate that the results of Thim's experiment cannot imply any contradiction in special relativity, which has been supported by a nearly uncountable number of experiments, including the transverse Doppler shifts predicted by Lorentz transformations for radiation sources moving with {\em constant velocities}.

Although there are no widely accepted exact rotational transformations, this does not mean that no exact rotational transformations exist. Indeed there are several in the literature, some of which are consistent with existing experimental tests.\cite{6,7} The lack of a consensus choice is more a reflection of the fact that establishing such a rotational transformation is not viewed as critical to advancing our understanding of physics. In this comment, we use one such exact transformation\cite{8} to re-analyze Thim's experiment, finding that his null result is fully consistent with the predictions of this transformation. Thus, rather than serving as a test of special relativity, Thim's experiment actually gives us clear clues about the nature of non-inertial frames in our universe and can help serve as a test of proposed space-time transformations between rotating frames. 

\section{Exact rotational space-time transformation}
The space-time transformations involving accelerations and rotations that we employ here have been discussed and derived in detail elsewhere in the literature\cite{6,8}. We simply give the result here. Suppose $F_{I}$, with coordinates $(w_{I},x_{I},y_I,z_I)$, is an inertial frame and $F(\Om)$, with coordinates $(w,x,y,z)$, is a frame that rotates with a constant angular velocity $\Om$ (to be defined more precisely below) with respect to $F_{I}$. The origins of both frames coincide at all times and, for reasons to be seen later, we use a Cartesian coordinate system in both frames. The time variable is expressed in units of length (where 1 meter is the amount of time it takes for a light signal in a vacuum to travel a distance of 1 meter in an inertial frame) to avoid complications associated with including the quantity $c$ in the transformation equations for non-inertial frames, in which the speed of light is not necessarily universal or constant\cite{6,9,10}. The exact rotational space-time transformation 
equations are\cite{6,8}
$$
	w_I =\g w,  \ \ \ \   x_I = \g[x \  cos(\Om w) - y \ sin(\Om w)],
$$	
\be
  	y_I = \g[x \ sin(\Om w) +  y \ cos(\Om w)],   \ \ \ \ \  z_I = z; 
\ee
%%%%%%%%11%%%%%%%%%2%%%1
$$
  \g = {1}/{\sqrt{1-\rh^2 \Om^2}}, \ \ \ \   \rho^2=x^2 + y^2;  
$$
and the inverse transformations are 
$$
	w= \frac{w_I}{\g}, \ \ \ \ \  x= \frac{1}{\g}[x_I \  cos(\Om_I w_I) + y_I  \ sin(\Om_I w_I)],
$$ 	
\be
  	y= \frac{1}{\g}[-x_I \ sin(\Om_I w_I) + y_I \ cos(\Om_I w_I)],  \ \ \ \ \  z= z_I;
\ee
%%%%%%%5.13%%%%%4%%%%6%%%2
$$
\g=\frac{1}{\sqrt{1- \Om_I^2(x^2_I +y^2_I)}}, \ \ \ \   \Om_I \equiv \frac{d \phi_I}{d w_I}=\frac{d \phi_I}{d w}\frac{d w}{d w_I}=\frac{\Om}{\g},  
$$
where $ \phi_I =\phi +\Om w$.
The quantity $\Om$ is the constant angular velocity of $F(\Om)$ with respect to $F_{I}$ as measured by observers in the $F(\Om)$ frame and $\Om_I$ is the angular velocity of $F(\Om)$ with respect to $F_{I}$ as measured by observers in the $F_{I}$ frame. The relationships between these angular velocities are 
\be
w_I  \Om_I  =w \Om , \ \ \ \   \rh_I \Om_I  =\rh \Om , \ \ \ \  \rh_I^2=x_I^2 +y_I^2.
\ee
%%%%%%%%%%%12%%%%%%3%%%%%5%%%3

The validity of these transformations has also been discussed elsewhere in detail\cite{6,8}. Here we make only two brief notes:

(A) Transformations (1) and (2) reduce to the Lorentz transformations in the appropriate limit. The transformation equations (1) are actually a special case of a more general set of transformation equations between an inertial frame and a non-inertial frame whose origin orbits the origin of the inertial frame at a radius $R$. The transformations between those two frames are\cite{6,8}
$$
	w_I = \g(w + \mbox{\boldmath$\rho\cdot\b$}), \ \ \ \ \  x_I = \g[x \  cos(\Om w) - (y-R) \ sin(\Om w)],$$ 	
\be
  	y_I = \g[x \ sin(\Om w) + ( y-R) \ cos(\Om w)],  \ \ \ \ \  z_I = z;
\ee
%%%%%%%5.10%%%%%%%%1%%%4
\be
\b = |\mbox{\boldmath$\Om$}\times  {\bf S}|=\Om \sqrt{x^2 +(y-R)^2}=\Om S<1, \ \ 
\ee
\be											
	 \mbox{\boldmath$\rho\cdot\b$}=x R\Om, \ \ \ \ \  \g=(1-\b^2)^{-1/2}.
\ee
In the limit of zero acceleration, i.e., when $R \to \infty$ and $\Om \to 0$ such that the product $R\Om = \b_o$ is a finite non-zero constant velocity, transformation (4) reduces to the Lorentz transformations
\be
 w_I=\g_o[w+x\b_o], \ \ \  x_I=\g_o[x+w\b_o], \ \ \  y_I=-\infty, \ \ \  z_I=z, 
 \ee
 %%%%%%%%%%%%%5
where one may shift the $y$-axis so that $y_I = y$. Thus, we see that Cartesian coordinates allow the exact rotational space-time transformation (4) to have this property, known as limiting Lorentz-Poincar\'e invariance\cite{8,11}. 
Furthermore, transformation (4) reduces to the classical rotational transformation in the case where $\b$ is small.

(B) Transformations (1) and (2) are consistent with all well-known experimental tests.  For example, consider the lifetime dilation of unstable particles traveling in a circular storage ring\cite{4}.  Furthermore, based on the rotational transformations of the covariant momentum vector, the expression for the energy of a particle, with rest mass $m$, traveling in a circle is  
\be
p_{I0} = \g m,
\ee
 in agreement with with the well-established results of high energy experiments performed in an inertial laboratory frame $F_I$.\cite{8}
 
 \section{Thim's experiment}

Thim's experiment\cite{1} can be analyzed using the rotational transformations of the covariant wave vector $k_{\mu}=(k_0, k_1,k_2,k_3)$ which, like the covariant momentum $p_\mu$, has the same transformation properties as the covariant coordinate differential vectors $dx_\mu$ and $dx_{I\mu}$.  The rotational transformations for the covariant coordinate differential vectors can be obtained from (1) with $dx_I^\mu=\e^{\mu\nu} dx_{I\nu}$ for inertial frames and $ dx^\mu=P^{\mu\nu} dx_{\nu}$ for rotational 
frames\cite{8,11}.\footnote{The non-vanishing components of $P_{\mu\nu}$ are given by  
$P_{00} = 1, P_{11} = -\g^2[1+2\g^2 \Om^2 x^2\\ - \g^4 \Om^4 x^2(w^2 -x^2 -y^2)], $ etc.
The contravariant metric tensors $P^{\mu\nu}$ are  $ P^{00} = \g^{-2}[1\\ -\Om^4 w^2 (x^2+y^2)],  \     
P^{11}=  -\g^{-2}[\g^{-2}(1-\Om^2 x^2) -2 \g^{-2} \Om^3 wxy +\Om^6 w^2  y^2 (x^2+y^2)], \ etc.$}  The transformation equations between the wave vector $k_{I\mu}$ measured in $F_I$ and the wave vector $k_{\mu}$ measured in the rotating frame $F(\Om)$ are\cite{6,8}
$$
k_{I0} = \g^{-1} (k_0 + \Om y k_1 -\Om x k_2),
$$
$$
k_{I1} =\left[-\g^{-2} \Om^2 wx_I\right]   k_0 + \g^{-2}\left[\g cos(\Om w) - \Om^2 x_I x - \Om^3 wx_I y\right]k_{1}
$$
%41%%%%%%%%%%%37%%36%%%%9%%3%%%%16%%%9
\be
+ \g^{-2}\left[  -\g sin(\Om w) - \Om^2 x_I y + \Om^{3} wx_I x\right] k_2,
\ee
$$
k_{I2} =\left[-\g^{-2} \Om^2 wy_I\right]   k_0 + \g^{-2}\left[\g sin(\Om w) - \Om^2 y_I x - \Om^3 wy_I y\right]k_{1}
$$
$$
+ \g^{-2}\left[ \g cos(\Om w) - \Om^2 y_I y + \Om^{3} w y_I x\right] k_2,$$
$$
k_{I3}=k_3, \ \ \ \ \   \g=\frac{1}{\sqrt{1-\rh^2 \Om^2}}.
$$

In Thim's experiment, the source is located on the axis of rotation of the $F(\Om)$ frame. Since all points on this axis of rotation are at rest in both the rotating and inertial frames, the space-time properties of both frames at all points on that axis are the same.  Thus, the frequency of the radiation from the source $f_I \equiv k_{I0}$ measured from $F_I$ is the same as the frequency of the radiation of the source at rest relative to $F(\Om)$ $f(rest)\equiv k_0(rest)$ measured from $F(\Om)$, 
\be
f_{I}=f_{I}(rest)=f(rest).
\ee
%%13%%15%%%%39%%%%%%%17%%%10
Consider one specific detector located at the rim of the rotating disk located at, say,  ${\rh}^i_a = (x_a,y_a,0)$.  Radiation with wave vector ${k_i}=(k_1, k_2,0)$ propagates from the center of the disk to this detector along the radius vector ${\rh^i}=(x,y,0)$, so that $k_1/k_2 =x/y.$   Thus, the first equation in (9) leads to
\be
f_{I}= \g^{-1} f, \ \ \ \ \   x k_2 = y k_1,  \ \ \ \   \g=\frac{1}{\sqrt{1-\rh_a^2 \Om^2}}.
\ee
%%%%%%%%14%%%16%%%%40%%%%%%%%%%%18%%%11
Combining (10) and (11), we obtain
\be
f =\g f_{I}= \g f(rest).
\ee
%%17%%%%%41%%%%%%%%%%%%%19%%%12
This result, which holds for each of the eight detectors, implies that observers at rest with respect to the detectors (i.e., observers at rest in the rotating frame) will measure a shift by a factor of $\g$ in the frequency of the radiation as a result of the orbiting motion of the detectors. Another way to think about this result is to see that according to the rotational transformations (1), clocks in a rotating frame located at a radius $\rho$ are slowed by a factor of $\g$, resulting in an increase in the detected frequency $f$ by a factor of $\g$.

However, result (12) actually does not contradict Thim's null result because in the experiment,\cite{1} the frequency of the radiation $f$ received by the orbiting detectors is not measured by observers in the rotating frame $F(\Om)$. Instead, the signal received by the detectors on the rotating disk is transferred through a second rotating disk to a stationary detector, where the frequency is measured by apparatus situated in the inertial laboratory $F_I$.  The process of transferring the signal back to the inertial laboratory frame is simply the reverse of the first, in which the frequency of the radiation was increased by a factor of $\g$. Thus, the frequency of the radiation $f_{I}(detector)$ received by the final detector, as measured in the laboratory, will be smaller than the frequency measured by the orbiting detectors by a factor of $\g$ because the clocks in the inertial frame run faster by the factor of $\g$, $f_{I}(detector)= f/\g$.  Combining these two processes leads to the result  
\be
 f_{I}(detector)=f (rest) =f_{I}.
\ee
%%%%%%%%%16%%%%%%18%%%%42%%%20%%%13
Thus, the rotational transformations (1) imply that when the frequency of the radiation $f_{I}(detector)$ received by the detector is measured by standard mixer and interferometer techniques in the inertial laboratory frame and compared with the frequency $f(rest)$ of the source, no shift should be measured, consistent with the results obtained by Thim.

\section{Discussion and conclusion}
The previous conclusion is independent of the angular velocity $\Om'$ of the second disk.  This can be seen by regarding the second disk as a second rotating frame $F(\Om')$.  The rotational transformations for $F_I$, $F(\Om)$ and $F(\Om')$ are
\be
w_I =\g w=\g' w',  \ \ etc.
\ee
%%%%%17%%%%%19%%%43%%%%21
\be
f_{I} = \g^{-1} f  = \g'^{-1} f', \ \ etc.
\ee
%%%%%%%%%%18%%%20%%%%%%%%%44%%%%22
$$
\g=\frac{1}{\sqrt{1-\rh_a^2 \Om^2}}, \ \ \ \ \ \   \g'=\frac{1}{\sqrt{1-\rh'^2 \Om'^2}}; 
$$
where $\rh_a$ and $\rh'$ are constant.  If the frequency of the radiation were to be measured by apparatus at rest in either the $F(\Om)$ (disk 1) or the $F(\Om')$ (disk 2) frame, it would differ from its original value (measured in the inertial laboratory frame). However, because this radiation is eventually transferred back and measured by a detector that is at rest in the same inertial laboratory frame as the source from which it was emitted, there will be no overall frequency shift. 

A new experimental test of the exact rotational transformations (1) is  to measure directly the frequency (12) of the signal received by the orbiting antenna on the rotating disk.  Assuming that the centrifugal effects on the apparatus are negligible, one can test the transverse frequency shift predicted in (12).  In fact, one could test for centrifugal effects by repeating the experiment with different angular velocities.

In conclusion, the Lorentz transformations cannot be used to analyze the results of Thim's experiment, involving orbiting detectors. Using the correct transformations for rotating non-inertial frames predicts that, as long as the signal received by the detectors is analyzed by apparatus at rest with respect to the inertial laboratory frame from which the radiation is initially emitted, no Doppler shift will be detected, consistent with the experimental observations.  To obtain a broader and more complete understanding of physics, it is desirable to generalize the physical framework to include non-inertial frames.   Inertial frames represent only limiting and idealistic cases and moreover, there is now strong evidence that the observable universe is expanding with a non-zero acceleration.  Therefore, rotational experiments of Davies-Jennison\cite{8} and Thim\cite{1} are important because they can reveal new principles in physics and increase our understanding of the physics of non-inertial frames. 

\bigskip

Note added in proof.   \ \ 
To avoid misunderstanding, we stress that it is incorrect to use the usual rotational coordinate transformations (i.e., $X=r \ cos(\theta -\om t),  \ Y= r \ sin(\theta - \om t),  \ Z=z,  \ cT=ct$; or equivalently, $ds^2=g_{\mu\nu} dx^\mu dx^\nu, g_{00}=1-\om^2 r^2/c^2,$ etc. See, for example, C Moller, The Theory of Relativity, p. 240) to discuss the precision experiments of Thim and Davis-Jennison mentioned in the present paper.\cite{1,8}  The reason is that this usual rotational coordinate transformation holds only approximately in the classical domain with small velocity $r \om << c$).  Furthermore,  this approximate rotational transformation is inconsistent with  the lifetime dilatation of muon decay in circular-orbital motion.\cite{4}  Therefore, one cannot use the result of this approximation rotational transformation, e.g., a rotating radius $r=\sqrt{x^2+y^2}$ does not contract, $x^2+y^2=X^2+Y^2$, to rule out the exact rotational transformations (1) or (4) in the present paper (which is consistent with the muon lifetime dilatation in circular motion and Thim's experiment) and to reject its logical consequence that a rotating radius contract.

 \newpage
%\begin{thebibliography} 
\bibliographystyle{unsrt}

\begin{thebibliography}{99}
\bibitem{1}H. W. Thim,  IEEE Trans.  Instrumentation and Measurement,  {\bf 52}, 1660 (2003).

\bibitem{2}A. Sfarti,  IEEE  Trans.  Instrumentation and Measurement, {\bf 59},  494 (2010).

\bibitem{3}H. W. Thim,  IEEE  Trans.  Instrumentation and Measurement, {\bf 59}, 495 (2010).

\bibitem{4}See, for example, F. J. M Farley, J. Bailey and E. Picasso, { Nature} {\bf 217}, 17 (1968). 

\bibitem{5}G. N. Pellegrini and A. R. Swift,  Am. J. Phys., {\bf 63},  694 (1995) and reference therein. 

\bibitem{6}J. P. Hsu and L. Hsu, in {\em A Broader View of Relativity, General Implications of Lorentz-Poincar\'e Invariance},  (Singapore, New Jersey, World Scientific (2nd Ed.), 2006), pp. 402-415 and pp. 267-318.

\bibitem{7}M. Nouri-Zonoz, H. Ramazani-Aval and R. Gharachahi,  arXiv 1208.1913v1, (2012), and reference therein.

\bibitem{8}L. Hsu and J. P. Hsu, Euro. Phys. J. Plus, {\bf 128}: 74 (2013).  (DOI  10.1140/epjp/i2013-13074-4). 

\bibitem{9}J. P. Hsu and L. Hsu,  Phys. Letts. A, {\bf 196}, 1 (1994). 

\bibitem{10}L. Hsu and J. P. Hsu, Eur. Phys. J. Plus, {\bf 127}, 11 (2012).  (DOI 10.1140/ epjp/i2012-12011-5).  

\bibitem{11}J. P. Hsu and L. Hsu, {\em Space-Time Symmetry and Quantum Yang-Mills Gravity},  (Singapore, New Jersey, World Scientific, 2013), Chapter 5.

  
\end{thebibliography}

\end{document}